\documentclass[twocolumn,floatfix,pra,showpacs,preprintnumbers,superscriptaddress,amsmath,amssymb,10pt,aps]{revtex4}
\usepackage{mathtools}
\usepackage{mathptmx}
\usepackage{subfigure}
\usepackage{psfrag,graphicx}
\usepackage{dcolumn}
\usepackage{adjustbox}
\usepackage{bm}
\usepackage{color}
\usepackage{latexsym}
\usepackage{epstopdf}
\usepackage{color}
\usepackage[english]{babel}
\usepackage{psfrag,graphicx}
\usepackage{subfigure}
\usepackage{amsmath}
\usepackage{amssymb}
\usepackage{amsfonts}
\usepackage{bm}
\usepackage{natbib}
\usepackage{epstopdf}
\usepackage{xcolor}
\usepackage{comment}
\usepackage[commandnameprefix=always]{changes}
\usepackage{appendix}

\definecolor{mygrey}{gray}{0.35}
\definecolor{myblue}{rgb}{0.2,0.2,0.8}
\definecolor{myzard}{cmyk}{0,0,0.05,0}
\definecolor{mywhite}{rgb}{1,1,1}
\definecolor{mywhite}{rgb}{1,1,1}
\definecolor{myred}{rgb}{1,0.,0.3}

\usepackage[colorlinks=true,citecolor=myblue,linkcolor=myred]{hyperref}

\def\ba{\begin{align}}
\def\enda{\end{align}}
\def\bi{\begin{itemize}}
\def\ei{\end{itemize}}

\def\be{\begin{equation}}
\def\ee{\end{equation}}
\def\bea{\begin{eqnarray}}
\def\eea{\end{eqnarray}}
\def\bse{\begin{subequations}}
\def\ese{\end{subequations}}

\definechangesauthor[name=bogomila, color=purple]{I}

\begin{document}
\title{Adiabatic Ramsey Interferometry for Measuring Weak Nonlinearities with Super-Heisenberg Precision}
\def\correspondingauthor{\footnote{Corresponding author: pivanov@phys.uni-sofia.bg}}
\author{Venelin P. Pavlov}
\affiliation{Center for Quantum Technologies, Department of Physics, St. Kliment Ohridski University of Sofia, James Bourchier 5 blvd, 1164 Sofia, Bulgaria}
\author{Bogomila S. Nikolova}
\affiliation{Center for Quantum Technologies, Department of Physics, St. Kliment Ohridski University of Sofia, James Bourchier 5 blvd, 1164 Sofia, Bulgaria}
\author{Peter A. Ivanov}
\affiliation{Center for Quantum Technologies, Department of Physics, St. Kliment Ohridski University of Sofia, James Bourchier 5 blvd, 1164 Sofia, Bulgaria}

\begin{abstract}
We propose an adiabatic Ramsey interferometry technique for detecting weak nonlinearities with trapped ions. The method relies on using the quantum Rabi model as a probe, which is sensitive to nonlinear symmetry-breaking perturbations. We show that the couplings which arise either from anharmonic terms of the trapping potential or due to higher order terms in the Coulomb interaction expansion can be efficiently estimated by measuring the spin state probabilities alone. We show that the spin signal is amplified by the mean-phonon excitations, which results in the estimation precision reaching the super-Heisenberg limit. Notably, achieving such high-precision estimation does not require specific entangled state preparation and can be reached even for initial thermal motion state. Furthermore, we show that the super-Heisenberg scaling can be observed even in the presence of weak spin-dephasing.  

\end{abstract}

\maketitle


\section{Introduction}
The measurement of weak signals is of key importance in various areas of modern physics. Quantum metrology is one of the leading branches of quantum technologies that uses quantum effects to improve measurement precision \cite{Degen2017,Pezze2018}. A well known example is Ramsey interferometry which aims to estimate an unknown phase $\varphi$ with $N$ uncorrelated two-state systems with statistical uncertainty $\delta\varphi\sim N^{-1/2}$ which is known as the standard quantum limit (SQL). Using an entangled state can further improve the sensitivity up to the so-called Heisenberg limit (HL), where the statistical uncertainty scales as $\delta\varphi\sim N^{-1}$ \cite{Wineland1992,Kitagawa1993}. Achieving such precision depends on both the initial state and the generator $\hat{G}$ of the unitary evolution $\hat{U}=e^{-i\varphi \hat{G}}$ that maps the unknown parameter to the quantum state. As in the case of a maximally entangled GHZ state and a linear generator, corresponding, for example, to frequency, magnetic- and electric-field estimation, the maximal precision is bounded by the HL \cite{Leibfried2004,Roos2006}. Crucially, HL is not fundamental, but can be overcome by utilizing a $k$-body interaction between the spins which can enhance the sensitivity to the super-Heisenberg (SH) limit where $\delta\varphi\sim N^{-k}$ with $k>1$ \cite{Boixo2007,Choi2008,Napolitano2011}, or by using critical quantum systems close to the quantum phase transition \cite{Rams2018,Gietka2023,Lyu2020}. Recently, a special class of spin-boson states was shown to also exhibit SH scaling of parameter estimation uncertainty \cite{Pavlov2025}. 

Another branch of quantum metrology considers parameter estimation of continuous variables \cite{Giovannetti2011,Giovannetti2006}. Here, the generator $\hat{G}$ maps the unknown parameter onto the state of the quantum oscillator, and the statistical uncertainty scales with the number of bosonic excitations rather than the number of spins. For example, the estimation of weak forces is bounded by SQL for initial coherent and squeezed states, and can be enhanced to HL using entangled states \cite{Munro2002,Maiwald2009,Ivanov2016}. The weak nonlinear coupling between the bosonic modes (photons or phonons) plays an important role in quantum optics \cite{Gerry2005}. For example, photon entanglement can be generated via the nonlinear process of spontaneous parametric down conversion (SPDC). Furthermore, the realization of a universal set of non-Gaussian quantum gates for continuous-variable quantum computation (CVQC) relies on the use of nonlinear bosonic coupling. The precise determination of such nonlinearity is an essential part of the realization of CVQC \cite{Lau2016} as well as pushing the fidelity of quantum gates toward the fault-tolerant threshold \cite{Sutherland2022}. Recently, an estimation of three-linear coupling between the vibrational modes of an ion crystal was discussed in \cite{Ivanov2022}. The parameter estimation is carried out either by detection of phonon probability distribution or by observation of Ramsey-type oscillations of the spin states. The estimation precision is bounded by the SQL and can be enhanced to HL by creating motion entangled states. An optimal design of the quantum state of the motional modes can further increase the sensitivity of the nonlinear parameter estimation to the SH limit \cite{Mahmoudi2024}. 
\begin{figure*}
\centering
    \adjustbox{margin*={-0.5cm -0.1cm}}{
        \includegraphics[width=0.7\textwidth]{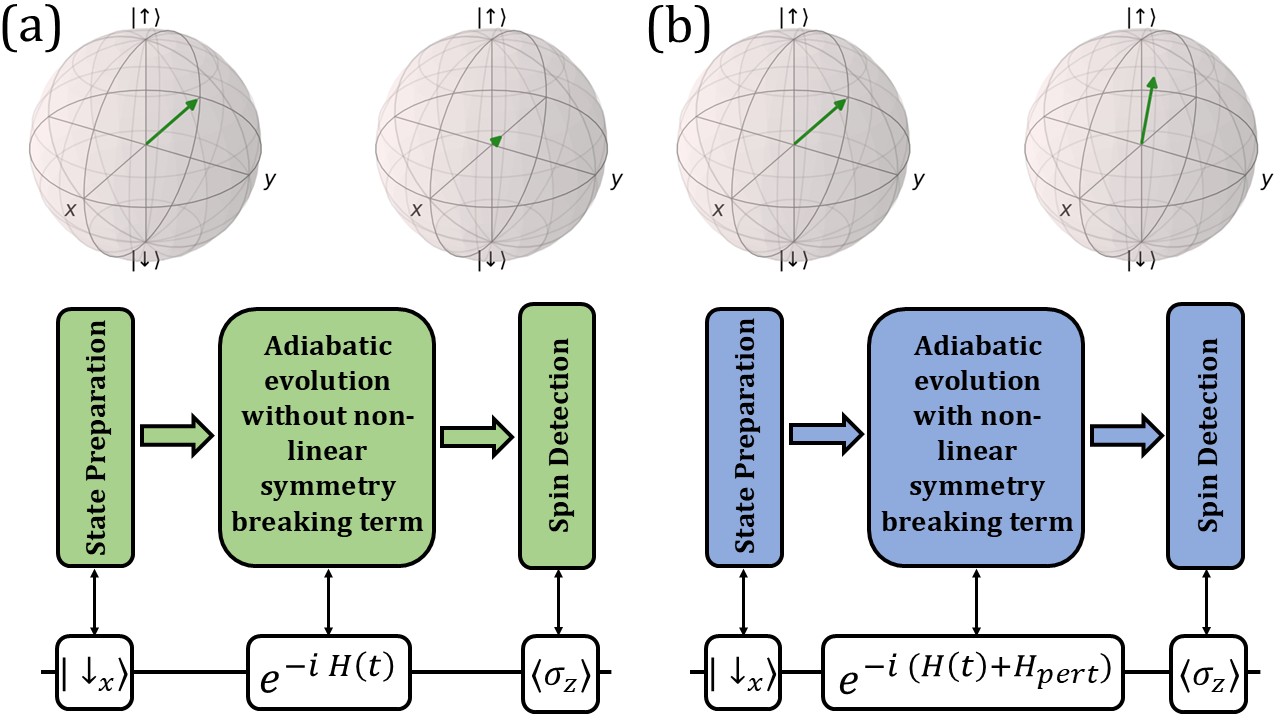}}
\caption{Nonlinear adiabatic Ramsey interferometer consists of initial state preparation, adiabatic evolution and spin detection. (a) Without symmetry-breaking term the adiabatic evolution creates a Schr\"odinger cat state between the spin and the motion mode with equal probability. The Bloch vector, initially pointing along the $x$-axis, points in the same direction at the end of the adiabatic evolution, which leads to $\langle\sigma_{z}\rangle=0$. (b) The presence of nonlinear symmetry breaking term leads to non-equally probable Schr\"odinger cat state at the end of the adiabatic transition. Consequently, the Bloch vector is rotated and $\langle\sigma_{z}\rangle\neq 0$.}
\label{fig1}
\end{figure*}

In this work, we discuss nonlinear adiabatic Ramsey interferometry for measuring nonlinear couplings with trapped ions. The technique relies on using the quantum Rabi model as a probe which is sensitive to nonlinear symmetry-breaking perturbation. This sensing technique was proposed in the context of the measurement of frequencies and weak forces \cite{Ivanov2013,Ivanov2015,Ivanov2020}. Here, we show that information about the nonlinear coupling can be efficiently mapped to the spin dynamics during the adiabatic evolution, and the parameter estimation is performed by measuring the spin probabilities alone. We consider two types of nonlinearity. In the first scenario, we consider the estimation of nonlinear coupling induced due to the presence of anharmonic terms in the trapping potential of the trapped ions \cite{Home2011,Johnson2011}. The effect of such nonlinearity becomes significant as ion trap sizes are reduced and its characterization is important for quantum information processing and quantum state engineering. Note that the effect is similar to SPDC of the $k$-th order, which was recently observed in a superconducting parametric-cavity system \cite{Chang2020}. 
Here, we show that, using our nonlinear Ramsey adiabatic technique, such nonlinearity can be measured by detecting spin probabilities with statistical uncertainty $\delta\varphi\sim \bar{n}^{-k/2}$ that approach the SH estimation sensitivity, with $\bar{n}$ being the average number of phonon excitations created during the adiabatic transition. In the second scenario, we discuss the parameter estimation of the three-linear coupling between the motion modes of the linear ion crystal. Such a nonlinearity can arise due to the higher-order terms in the Coulomb interaction expansion which cause coupling between the motion modes of the trapped ions \cite{Roos2008,Ding2017_1,Ding2017_2,Maslennikov2019}. We show that parameter estimation can be carried out by spin detection with precision bounded by the SH limit. 

Crucial advantages of our adiabatic sensing technique are (i) measurement of single spin observable and improvement of the estimation precision up to SH limit in terms of the number of phonon excitations rather than with the number of particles, (ii) the average number of phonon excitations is controlled either by the spin-phonon coupling and motion squeezing during the adiabatic transition, or by the initial coherent state amplitudes, (iii) the sensing technique does not rely on specific initial entangled state preparation and can be applied even for a thermal phonon state and outside the Lamb-Dicke limit, and (iv) the SH limit of the parameter estimation precision can be observed even in the presence of weak spin dephasing.

The paper is organized as follows: In Sect. \ref{SM Rabi} we introduce the quantum Rabi model with additional motion squeezing term as a quantum probe sensitive to nonlinear symmetry-breaking perturbations. In Sect. \ref{AT} we discuss the effect of the nonlinearity onto the spin probabilities during the adiabatic evolution. We show that the weak nonlinearity rotates the Bloch vector during the adiabatic transition, and thus the parameter estimation is carried out by detecting the spin probabilities. In Sections \ref{SM} and \ref{MM} we show that estimation of the coupling of single-mode and multi-mode nonlinearity can be achieved by detecting the spin probability alone, and the precision is bounded by the SH limit. In Sect. \ref{EC} we discuss some experimental imperfections that can limit the estimation precision. Finally, we present the conclusion in Sect. \ref{C}.  

\section{Quantum Rabi model}\label{SM Rabi}
We consider the quantum Rabi model in the presence of motion squeezing. The Hamiltonian is given by
\begin{eqnarray}
\hat{H}(t)&=&\hbar\omega \hat{a}^{\dag}\hat{a}+\frac{\hbar\Omega(t)}{2}\sigma_{x}+
\hbar g\sigma_{z}(\hat{a}^{\dag}+\hat{a})\notag\\
&&+\hbar\xi(\hat{a}^{\dag2}e^{i\theta}+\hat{a}^{2}e^{-i\theta}),\label{model}
\end{eqnarray}
where $\hat{a}^{\dag}$ and $\hat{a}$ are the creation and annihilation operators of the phonon excitation with frequency $\omega$, and $\Omega(t)$ is the time-dependent Rabi frequency of the two-level system with $\sigma_{q}$ ($q=x,y,z$) being the Pauli matrices. The coupling $g$ quantifies the strength of the dipolar spin-phonon interaction. The last term in (\ref{model}) describes the motion squeezing with strength $\xi$ and phase $\theta$ which we set to $\theta=\pi$. This term can be realized, for example, through the application of an additional time-varying electric field or by modulating the confining potential of the ion chain \cite{Burd2019,Burd2021}. 
Although we consider the trapped ion realization of the model here \cite{Lv2018}, other quantum-optical platforms are also suitable for its realization including, for example, circuit quantum electrodynamics chip \cite{Langford2017} and the superconducting circuit embedded in the cQED setup \cite{Braumuller2017}.

The quantum Rabi model predicts exotic phenomena such as finite size quantum phase transition \cite{Ashhab2013,Bakemeier2012,Hwang2015}, quantum chaos \cite{Sun2020,Kirkova2022}, and ground-state entanglement \cite{Pedernales2015}.
The Hamiltonian (\ref{model}) has a parity symmetry generated by the operator $\hat{\Pi}=\hat{a}^{\dag}\hat{a}+\frac{1}{2}(\sigma_{z}+1)$ such that $[e^{i\pi\hat{\Pi}},\hat{H}]=0$. In general, the model (\ref{model}) is not exactly solvable, except in a few limits. 
\begin{figure}[!htbp]
\hspace*{-0.35cm}
\includegraphics[width=0.5\textwidth]{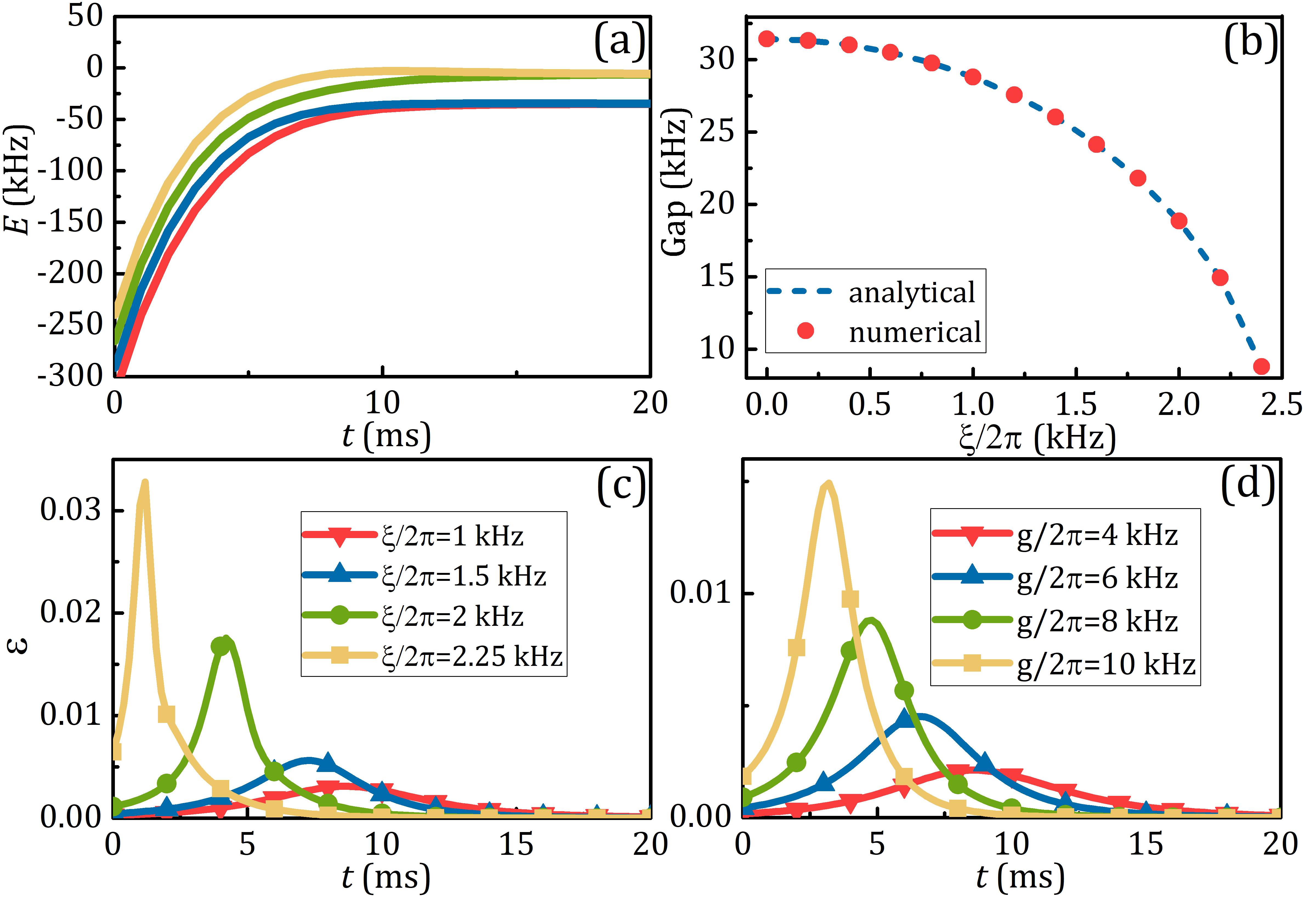}
\caption{(a) The first four eigenenergies $E_n$ of $\hat{H}(t)$ as a function of time $t$ for $\xi/2\pi$ = 1.0 kHz.  (b) The gap between the energies of the degenerate ground state and the first degenerate excited state of $\hat{H}(t)$ at $t_{f}$ for different values of the motion squeezing strength $\xi$. The dashed line shows $\tilde{\omega}$. The parameters are set to $\omega/ 2 \pi= 5 $ kHz, $\Omega/ 2 \pi= 100$ kHz, $g/ 2 \pi = 4$ kHz, $\tau = 3.5$ ms. (c) Adiabatic parameter $\varepsilon$ as a function of time for different values of $\xi$. Here $\Omega/ 2 \pi= 150$ kHz and $f_{3}/2\pi= 0.5$ Hz. (d) Adiabatic parameter $\varepsilon$ as a function of time for different values of the spin-phonon coupling $g$ and for $\xi=0$. }
\label{fig2}
\end{figure}

\subsection{Limit Case \texorpdfstring{$\Omega\gg g$}{Omega >> g}}
First, we study the limit $\Omega\rightarrow\infty$ which is equivalent to setting $g=0$ in (\ref{model}). In that case, the phonon and spin degrees of freedom are decoupled and can be diagonalized independently. The motion Hamiltonian $\hat{H}_{\rm b}=\hbar\omega \hat{a}^{\dag}\hat{a}-\hbar\xi(\hat{a}^{\dag 2}+\hat{a}^{2})$ can be brought in a diagonal form after an application of the transformation $\hat{\tilde{H}}_{\rm b}=\hat{S}^{\dag}(r)\hat{H}_{\rm b}\hat{S}(r)$, where $\hat{S}(r)=e^{\frac{r}{2}(\hat{a}^{\dag2}-\hat{a})}$
is the squeezing operator with amplitude 
\begin{equation}
r=\frac{1}{4}\ln\left(\frac{1+2\xi/\omega}{1-2\xi/\omega}\right).\label{SA}
\end{equation}
We have
\begin{equation}
\hat{\tilde{H}}_{\rm b}=\hbar\tilde{\omega}\hat{a}^{\dag}\hat{a}+E,
\end{equation}
where $\tilde{\omega}=\omega\sqrt{1-(2\xi/\omega)^{2}}$ and $E=\hbar\omega\sinh^{2}(r)-\hbar\xi\sinh(2r)$. Therefore, including the spin component, the non-degenerate eigenenergies are $E_{n,\pm}=n\tilde{\omega}\pm(\Omega/2)+E$ and their corresponding eigenvectors are
\begin{equation}
|\varphi_{n,\pm}\rangle=\hat{S}(r)|n\rangle|\pm\rangle,
\end{equation}
where $|n\rangle$ ($n=0,1,2,\ldots$) is the Fock state of the phonon mode excitation and $\sigma_{x}|\pm\rangle=\pm|\pm\rangle$.

\subsection{Limit Case \texorpdfstring{$\Omega=0$}{Omega = 0}}
We now study the limit $\Omega(t)=0$. In that case, the Hamiltonian $\hat{H}_{\Omega=0}$ is diagonal in the spin basis and the remaining phonon part is quadratic, so it can be diagonalized. First, a squeezing transformation is applied, $\hat{S}^{\dag}(r)\hat{H}_{\Omega=0}\hat{S}(r)$ with $r$ given by \eqref{SA}, and a subsequent displacement transformation $\hat{\tilde{H}}_{\Omega=0}=\hat{D}^{\dag}(\beta)\hat{S}^{\dag}(r)\hat{H}_{\Omega=0}\hat{S}(r)\hat{D}(\beta)$, with $\hat{D}(\beta)=e^{-\sigma_{z}\beta(\hat{a}^{\dag}-\hat{a})}$ being the spin-dependent displacement operator with amplitude $\beta=-(ge^{r}/\tilde{\omega})$. We obtain $\hat{\tilde{H}}_{\Omega=0}=\hbar\tilde{\omega}\hat{a}^{\dag}\hat{a}+e_1$, where $e_1=-\hbar (g^{2}/\tilde{\omega})e^{2r}+e$. Hence, the eigenenergies of $\hat{H}_{\Omega=0}$ are doubly degenerate with the corresponding eigenvectors: 
\begin{eqnarray}
|\psi_{\uparrow,n}\rangle=\hat{S}(r)\hat{D}(\beta)|n\rangle\left|\uparrow\right\rangle,\quad 
|\psi_{\downarrow,n}\rangle=\hat{S}(r)\hat{D}(\beta)|n\rangle\left|\downarrow\right\rangle.\label{state}
\end{eqnarray}
Here, $\left|\uparrow\right\rangle$ and $\left|\downarrow\right\rangle$ are the eigenstates of the $\sigma_{z}$ Pauli matrix.

Let us now consider the term $\hat{H}_{\rm spin}=(\hbar\Omega/2)\sigma_{x}$ by treating it as a perturbation. The effect of $\hat{H}_{\rm spin}$ is to lift the degeneracy of the Hamiltonian eigenspectrum $\hat{H}_{\Omega=0}$. Using degenerate perturbation theory, one can evaluate the energy splitting (gap) between each doubly degenerate manifold. The energy gap is $\Delta_{n}=\Omega\langle\psi_{\uparrow,n}|\sigma_{x}|\psi_{\downarrow,n}\rangle$ and, using (\ref{state}), we obtain
\begin{equation}
\Delta_{n}=\Omega e^{-2\beta^{2}}L_{n}(4\beta^{2}),
\end{equation}
where $L_{n}(x)$ is the Laguerre polynomial.

\section{Adiabatic Transition}\label{AT}
The sketch of the nonlinear adiabatic Ramsey interferometry is shown in Fig.\ref{fig1}. We assume that an additional symmetry-breaking term is applied and the total Hamiltonian then becomes
\begin{equation}
\hat{H}_{\rm T}(t)=\hat{H}(t)+\hat{H}_{\rm pert}.
\end{equation}
The system is initially prepared in the ground state of Hamiltonian (\ref{model}). For $\Omega(0)\gg g$ the ground state is unique and is given by $|\psi(0)\rangle=|-\rangle|r\rangle$ where $|r\rangle$ is a phonon squeezed state. Then, we lower the Rabi frequency adiabatically so that, at $t_{f}$, we have $\Omega(t_f)\ll g $. The corresponding ground state is a Schr\"odinger cat state, 
\begin{equation}
|\psi(t_f)\rangle=c_{\uparrow}(t_{f})|\psi_{\uparrow,0}\rangle+c_{\downarrow}(t_{f})|\psi_{\downarrow,0}\rangle.
\end{equation}
In the absence of the symmetry-breaking term $\hat{H}_{\rm pert}=0$, the probabilities are equal $|c_{\uparrow}(t)|^{2}=|c_{\downarrow}(t)|^{2}=\frac{1}{2}$. Consider the Bloch vector $\vec{B}(t)=[\langle\sigma_{x}(t)\rangle,\langle\sigma_{y}(t)\rangle,\langle\sigma_{z}(t)\rangle]^{T}$ that is  initially aligned along the $x$-axis, namely $\langle\sigma_{x}(0)\rangle=-1$ and $\langle\sigma_{y}(0)\rangle=\langle\sigma_{z}(0)\rangle=0$. During the adiabatic evolution, spin-phonon entanglement is created, which reduces the length of the Bloch vector such that we now have $\vec{B}(t_{f})=[e^{-\frac{1}{2}\beta^{2}},0,0]^{T}$, see Fig. \ref{fig1}(a). 
\begin{figure}[!htbp]
\hspace*{-0.1cm}
\includegraphics[width=0.49\textwidth]{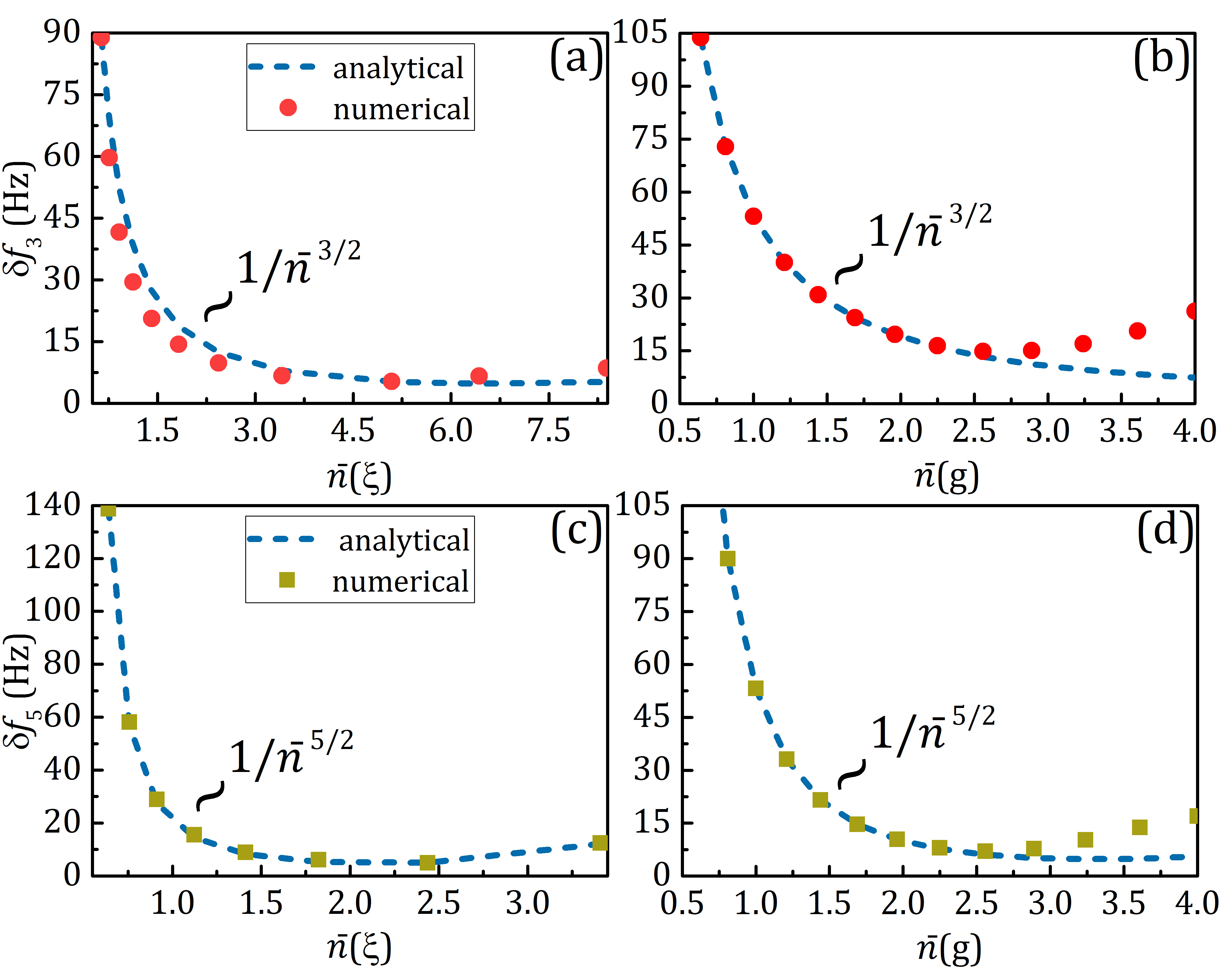}
\caption{(a) Parameter estimation precision for $f_3/2\pi$ = 0.5 Hz as a function of the average number of phonon excitations $\bar{n}(\xi)$ for $\tau=3.5$ ms. The dashed line shows the analytical formula (\ref{errorH3}). (b) Parameter estimation precision for $f_3$ with respect to $\bar{n}(g)$ for $\tau=3.0$ ms, $\xi= 0$ and $\Omega/2\pi= 150$ kHz. (c) Parameter estimation precision for $f_5/2\pi$ = 0.5 Hz as a function of the average number of phonon excitations $\bar{n}(\xi)$ for $\tau=3.5$ ms. The dashed line shows the analytical formula (\ref{errorH5}) (d) Parameter estimation precision for $f_5$ with respect to $\bar{n}(g)$ for $\tau=3.0$ ms, $\xi= 0$ and $\Omega/2\pi= 150$ kHz.}
\label{fig3}
\end{figure}

Crucially, for $\hat{H}_{\rm pert}\neq 0$, the parity symmetry of Hamiltonian (\ref{model}) is broken, which leads to $|c_{\uparrow}(t)|^{2}\neq |c_{\downarrow}(t)|^{2}$. In order to evaluate the effect of the symmetry-breaking term on the ground state, we represent the Hamiltonian (\ref{model}) within the ground state multiplet. As long as the term $\hat{H}_{\rm pert}$ does not couple excited states with the ground state manifold, the effective description of the probability amplitudes $c_{s}(t)$ ($s=\uparrow,\downarrow$) is reduced to a two-state problem. Then we have \cite{Ivanov2013}:
\begin{eqnarray}
&&i\frac{d}{dt}c_{\uparrow}(t)=-\alpha c_{\uparrow}(t)+\frac{\Delta_{0}(t)}{2}c_{\downarrow}(t),\notag\\
&&i\frac{d}{dt}c_{\downarrow}(t)=\alpha c_{\downarrow}(t)+\frac{\Delta_{0}(t)}{2}c_{\uparrow}(t),\label{TSP}
\end{eqnarray}
where 
\begin{equation}
\alpha=\langle\psi_{\downarrow,0}|\hat{H}_{\rm pert}|\psi_{\downarrow,0}\rangle\label{ME}
\end{equation}
and $\Delta_{0}(t)=\Omega(t)e^{-2\beta^{2}}$ is the ground-state energy gap. The effect of the symmetry-breaking term is to rotate the Bloch vector which becomes, see Fig. \ref{fig1}(b),
\begin{eqnarray}
&&\langle\sigma_{x}(t_{f})\rangle=e^{-\frac{1}{2}\beta^{2}}\Re (c_{\uparrow}(t_{f})c_{\downarrow}(t_{f})),\notag\\
&&\langle\sigma_{y}(t_{f})\rangle=e^{-\frac{1}{2}\beta^{2}}\Im (c_{\uparrow}(t_{f})c_{\downarrow}(t_{f})),\notag\\
&&\langle\sigma_{z}(t_{f})\rangle=|c_{\uparrow}(t_{f})|^{2}-|c_{\downarrow}(t_{f})|^{2}.\label{bloch}
\end{eqnarray}

Consider a time dependent Rabi frequency of the form $\Omega(t)=\Omega_{0}e^{-t/\tau}$. The symmetry-breaking term is \eqref{HD}, and the two-state problem (\ref{TSP}) is reduced to the Demkov model. The solution is then given by:
\begin{eqnarray}
&&c_{\uparrow}(z)=\frac{\pi}{2\sqrt{2}}\frac{e^{-\frac{t}{2\tau}}x}{\cosh(\pi\alpha\tau)}
\{J_{\nu}(z)(J_{1-\nu}(x)-iJ_{-\nu}(x))\notag\\
&&\quad\quad\quad+J_{-\nu}(z)(J_{\nu-1}(x)+iJ_{\nu}(x))\},\notag\\
&&c_{\downarrow}(z)=-\frac{\pi}{2\sqrt{2}}\frac{e^{-\frac{t}{2\tau}}x}{\cosh(\pi\alpha\tau)}\{J_{\nu-1}(z)(J_{-\nu}(x)+iJ_{1-\nu}(x))\notag\\
&&\quad\quad\quad+J_{1-\nu}(z)(J_{\nu}(x)-iJ_{\nu-1}(x)).
\end{eqnarray}
Here $J_{\nu}(z)$ is a Bessel function of the first kind, where $z=x e^{-\frac{t}{\tau}}$, $x=\frac{\Omega_{0}\tau}{2} e^{-2\beta^{2}}$ and $\nu=\frac{1}{2}+i\alpha\tau$. Using the asymptotics $J_{\nu}(z)\sim \frac{(z/2)^{\nu}}{\Gamma(1+\nu)}$ for $t\gg\tau$, with $\Gamma(1+z)$ being the Gamma function, and $J_{\nu}(z)\sim\sqrt{\frac{2}{\pi z}}\cos(z-\frac{\nu\pi}{2}-\frac{\pi}{4})$ for $x\gg|\nu^{2}-1/4|$, we obtain the following expressions, \cite{Ivanov2013}:
\begin{eqnarray}
&&c_{\uparrow}(z)\approx e^{ix}\sqrt{\frac{\pi}{2}}\frac{(z/2)^{-i\alpha\tau}}{\cosh(\pi\alpha\tau)}\frac{e^{\frac{\pi\alpha\tau}{2}}}{\Gamma(\frac{1}{2}-i\alpha\tau)},\notag\\
&&c_{\downarrow}(z)\approx -e^{ix}\sqrt{\frac{\pi}{2}}\frac{(z/2)^{i\alpha\tau}}{\cosh(\pi\alpha\tau)}\frac{e^{-\frac{\pi\alpha\tau}{2}}}{\Gamma(\frac{1}{2}+i\alpha\tau)}.\label{PA}
\end{eqnarray}

\subsection{Adiabatic condition}
The first four energies of the Hamiltonian (\ref{model}) are shown in Fig. \ref{fig2}(a). In both limits $\Omega\ll g$ and $\Omega\gg g$ the energy gap to the nearest excited energy is $\tilde{\omega}$. For large squeezing the gap decreases, and thus, for time-dependent Rabi frequency $\Omega(t)=\Omega_{0}f(t/\tau)$, the adiabatic condition for the system to follow the ground state requires $\tau\gg\tilde{\omega}^{-1}$. In order to characterize the non-adiabatic transitions, we define the parameter $\epsilon=|\langle\psi_{\rm exc}|\frac{d}{dt}|\psi_{\rm g}\rangle/\Delta_{\rm gap}|$. The adiabatic condition requires that $\epsilon\ll 1$. Here $|\psi_{\rm exc}(t)\rangle$ and $|\psi_{g}(t)\rangle$ are the instantaneous excited and ground states which are separated through the energy gap $\Delta_{\rm gap}(t)$. We see that the non-adiabatic transitions increase with $\xi$ and $g$ and hence, to fulfill the adiabatic condition, one needs a longer interaction time and larger $\tau$.

\subsection{Spin observable}
In order to perform the parameter estimation, we can measure one of the components of the Bloch vector (\ref{bloch}). Since the components $\langle\sigma_{x,y}(t_{f})\rangle$ decrease with increasing amplitude $\beta$, we choose the observable to be $\langle\sigma_{z}(t_{f})\rangle$. For a general symmetry-breaking term $\hat{H}_{\rm pert}=\lambda f(\hat{a}^{\dag},\hat{a})$, the signal $\langle \sigma_{z}(t_{f})\rangle$ becomes sensitive to $\lambda$, and thereby it can be estimated by detecting the spin probabilities after the adiabatic transition. Using (\ref{PA}) the signal becomes, \cite{Ivanov2013}:
\begin{equation}
\langle \sigma_{z}(t_{f})\rangle=
\tanh(\pi\alpha\tau).
\end{equation}
As a figure of merit for the sensitivity, we use the error propagation formula 
\begin{equation}
\delta\lambda=\frac{\sqrt{1-\langle\sigma_{z}(t_{f})\rangle^{2}}}{\left|\frac{\partial\langle\sigma_{z}(t_{f})\rangle}{\partial\lambda}\right|}.
\end{equation}
For comparison of the optimality of parameter estimation we also use the quantum Fisher information (QFI), \cite{Paris2009}:
\begin{figure}[!htbp]
\hspace*{-0.2cm}
\includegraphics[width=0.495\textwidth]{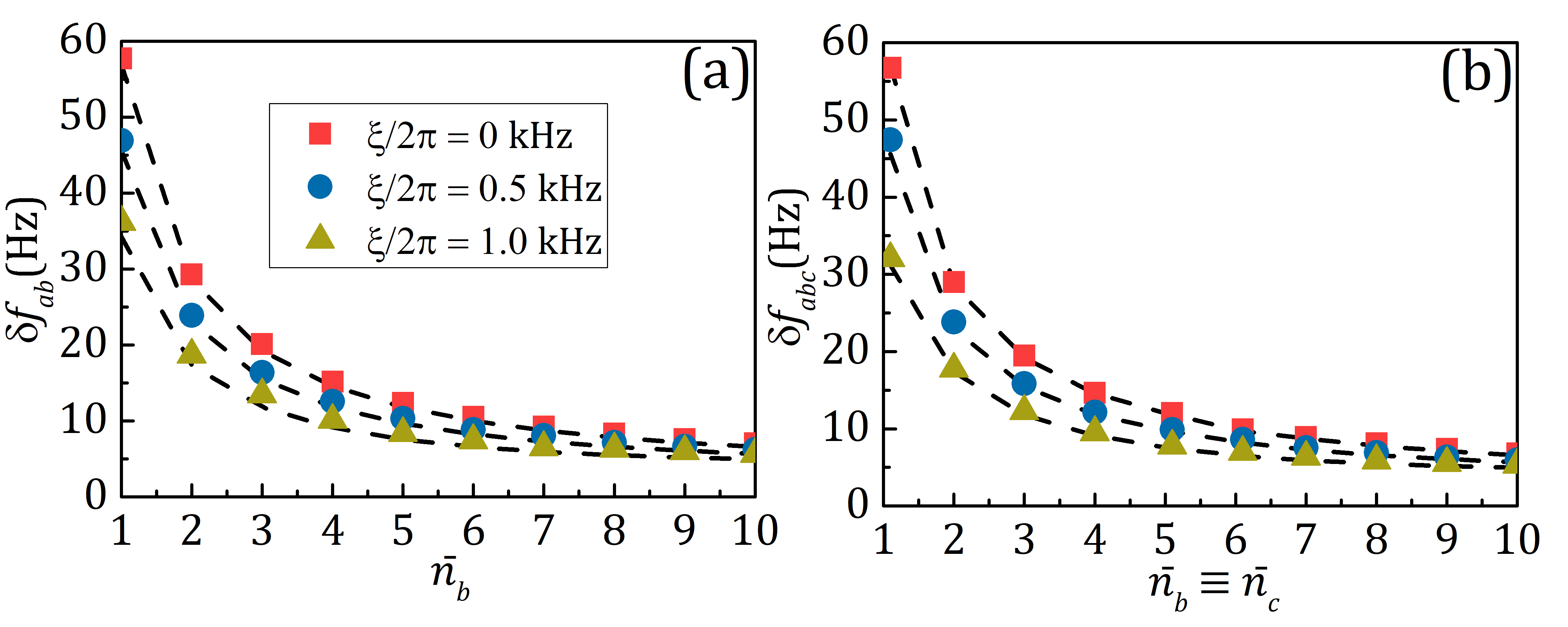}
\caption{(a) Parameter estimation precision for $f_{ab}/2\pi$ = 0.5 Hz as a function of the average number of phonon excitations $\bar{n}_b$ for $\tau=3.5$ ms and $\Omega/2\pi$ = 100 kHz. The dashed line shows the analytical formula (\ref{errfab}). (b) Parameter estimation precision for $f_{abc}/2\pi$ = 0.5 Hz with respect to $\bar{n}_b\equiv\bar{n}_c$. The dashed line shows the analytical formula (\ref{errfabc})}
\label{fig4}
\end{figure}
\begin{equation}
F_{Q}(\lambda)=4(\langle\partial_{\lambda}\psi(t)|\partial_{\lambda}\psi(t)\rangle-|\langle\psi(t)|\partial_{\lambda}\psi(t)\rangle|^{2}).
\end{equation}
The ultimate precision in parameter estimation is quantified by the Cramer-Rao bound $(\delta\lambda)_{\rm opt}=1/\sqrt{F_{Q}(\lambda)}$, such that $(\delta\lambda)_{\rm opt}\leq\delta\lambda$, where the equality is maintained as long as the measurement basis is optimal. We find
\begin{eqnarray}
F_{Q}(\lambda)&=&(\partial_{\lambda}\alpha)^{2}\tau^{2}{\rm sech}^{2}(\pi\alpha\tau)\{\pi^{2}+\ln^{2}(4)+4\ln(z/4)\ln(z)\notag\\
&&+2\Re\psi_{0}(\nu)(\ln(16)-4\ln(z)+2\Re\psi_{0}(\nu)\},
\end{eqnarray}
where $\psi_{n}(\nu)$ is the polygamma function of the $n$-th order. For $t\gg\tau$ one can approximate the QFI so that we obtain
\begin{equation}
F_{Q}(\lambda)\approx 4 (\partial_{\lambda}\alpha)^{2} t^{2}{\rm sech}^{2}(\pi\alpha\tau)\label{QFI}.
\end{equation}
In the following, we show that the signal is amplified by the mean phonon excitation, which improves the statistical uncertainty to the SH limit. 
 
\section{Single-mode nonlinearity}\label{SM}
We begin by considering a symmetry-breaking term of the following form:
\begin{equation}
\hat{H}_{\rm pert}=\hbar f_{k}(\hat{a}^{\dag k}+\hat{a}^k), \label{HD}
\end{equation}
which is the so-called spontaneous parametric down-conversion of the $k$-th order \cite{Chang2020}. Here $\lambda=f_{k}$ is the parameter we wish to estimate. For an odd number $k$, the term $\hat{H}_{\rm pert}$ breaks the parity symmetry $\hat{\Pi}\hat{H}_{\rm pert}\hat{\Pi}^{\dag}=-\hat{H}_{\rm pert}$. In this case, the matrix element (\ref{ME}) is given by:
\begin{equation}
\begin{aligned}
\alpha = f_k \bigg\{ e^{kr}{\beta}^k
+ \sum_{n=0}^k \binom{k}{n}
  {\cosh(r)}^{k-n}{\sinh(r)}^{n} \\
\times (-1)^{k-n}(-\beta)^{k-2n}
   U(-n,1-2n+k,-{\beta}^2) \bigg\},\label{alpha}
\end{aligned}
\end{equation}
where $U(a,b,z)$ is the Tricomi confluent hypergeometric function.

\subsection{$k=3$}
In this subsection, we will focus on the cubic nonlinearity which may arise as a result of anharmonic terms in the trapping potential of trapped ions \cite{Home2011}, (see Appendix \ref{PR}). This nonlinear term may become significant as the trap size decreases.  Recently, such a cubic interaction Hamiltonian was experimentally observed with a flux‑pumped superconducting cavity \cite{Chang2020}.

Using (\ref{alpha}) we find that the matrix element is given by
\begin{equation}
\alpha=f_{3}(2\beta^{3}e^{3r}+3\beta e^{r}\sinh(2r)).
\end{equation}
It is convenient to express $\alpha$ in terms of the average number of phonon excitations that are created at the end of the adiabatic evolution, namely $\bar{n}=\langle\psi(t_{f})|\hat{a}^{\dag}\hat{a}|\psi(t_{f})\rangle$. We have
\begin{equation}
\bar{n}=\frac{g^{2}}{\omega^{2}}\frac{1}{\left(1-\frac{2\xi}{\omega}\right)^{2}}+\frac{1}{2}\frac{1}{\sqrt{1-\left(\frac{2\xi}{\omega}\right)^{2}}}-\frac{1}{2}.
\end{equation}
For squeezing $2\xi<\omega$, one can approximate $\bar{n}\approx (g/\omega)^{2}(1-2\xi/\omega)^{-2}$ and $\alpha\approx 2f_{3}\bar{n}^{3/2}$. The signal becomes $\langle\sigma_{z}(t_{f})\rangle=\tanh(2\pi\tau f_{3}\bar{n}^{3/2})$ and the parameter estimation precision is given by
\begin{equation}
\delta f_{3}=\frac{1}{2\pi\tau \bar{n}^{3/2}}\cosh(2\pi\tau f_{3}\bar{n}^{3/2}).\label{errorH3}
\end{equation}
Therefore, the signal is amplified with the average number of phonon excitations, which improves the statistical uncertainty up to the SH limit. 

In Figs. \ref{fig3}(a) and \ref{fig3}(b) we show the exact result for the error propagation formula as a function of the average number of phonon excitations which are created during the adiabatic evolution. We can control $\bar{n}$ either by changing the squeezing $\xi$ or the spin-phonon coupling $g$. We see in both cases that the statistical uncertainty shows SH limit. Decreasing $\tau$ leads to stronger non-adiabatic transitions which spoil the parameter estimation for large $\bar{n}$, as shown in \ref{fig3}(b). We note that increasing $\bar{n}$ further decreases the estimation precision, since the factor $\cosh(2\pi\tau f_{3}\bar{n}^{3/2})$ in (\ref{errorH3}) increases for $2\pi\tau f_{3}\bar{n}^{3/2}>1$. Therefore, a significant enhancement of parameter estimation can be achieved as long as $2\pi\tau f_{3}\bar{n}^{3/2}<1$.

Finally, we emphasize that, using Eq. (\ref{QFI}) the optimal statistical uncertainty is $(\delta f_{3})_{\rm opt}=\cosh(2\pi\tau f_{3}\bar{n}^{3/2})/4t\bar{n}^{3/2}$ and, hence, the estimation precision is improved by $t$ rather than $\tau$ ($t\gg\tau$) unlike the dependence in Eq. (\ref{errorH3}). However, achieving such high precision requires measurements not only of the spin component but also of the motion degrees of freedom, which inevitably leads to a more complex measurement protocol. 

\subsection{$k=5$}

Using (\ref{alpha}) we find that the matrix element is given by
\begin{equation}
\alpha=f_{5}\{2\beta^{5}e^{5r}+5\beta^3 (e^{5r}-e^{r})+\frac{15}{4}\beta(e^{-3r}+e^{5r}-2e^r)\}.
\end{equation}
In order to emphasize the SH scaling we consider $\xi=0$ such that we have $\alpha=2f_{5}\bar{n}^{5/2}$ and the statistical uncertainty becomes, see Fig. \ref{fig2}(d):
\begin{equation}
\delta f_{5}=\frac{1}{2\pi\tau \bar{n}^{5/2}}\cosh(2\pi\tau f_{3}\bar{n}^{5/2}).\label{errorH5}
\end{equation}

\begin{figure}[!htbp]
\hspace*{-0.1cm}
\includegraphics[width=0.48\textwidth]{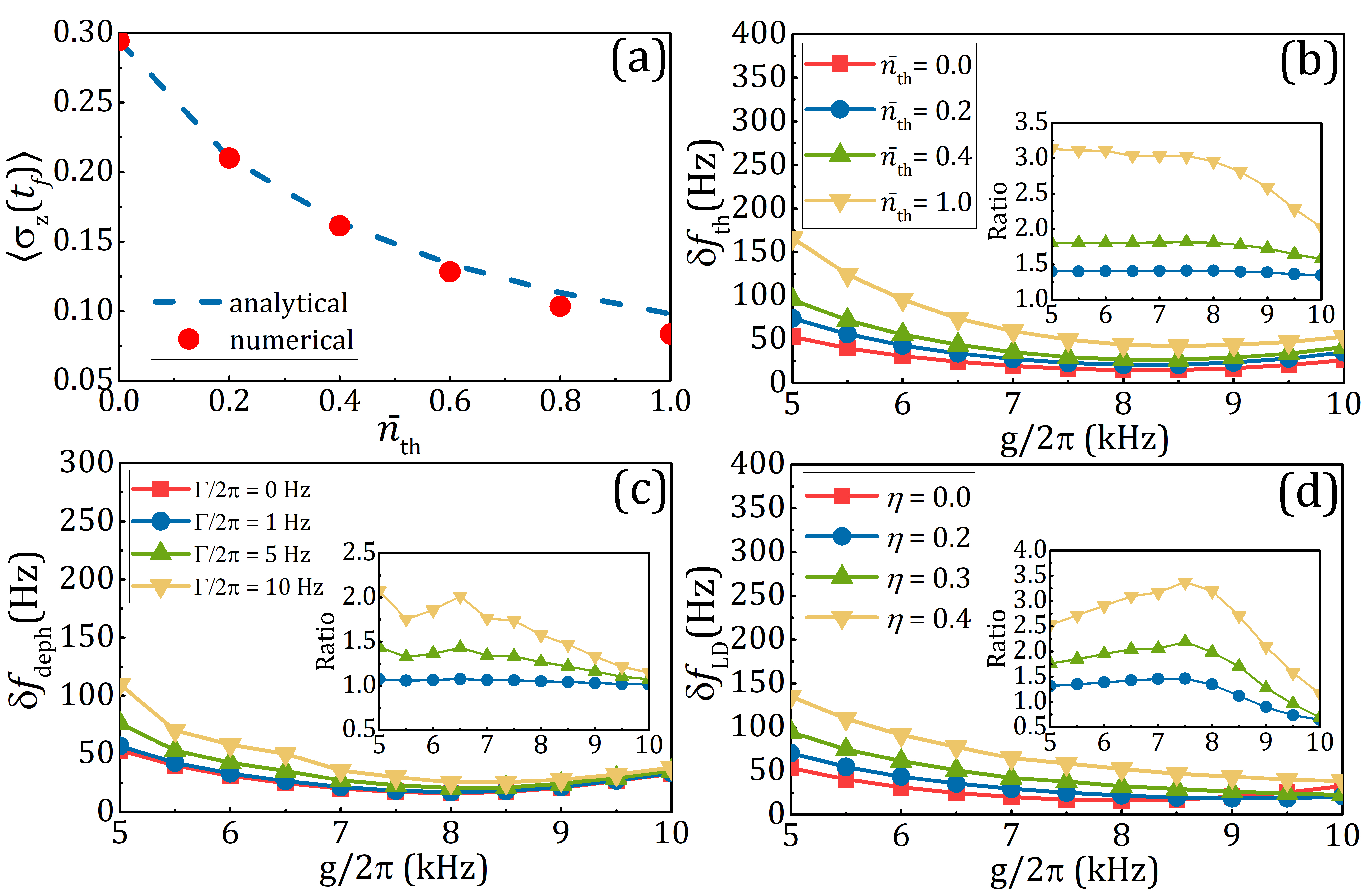}
\caption{(a) The spin observable $\langle\sigma_z\rangle$ as a function of $\bar{n}_{\rm th}$ at $t=t_f$. The numerical result is compared with the analytical formula (\ref{signal_th}). (b) Parameter estimation precision for initial thermal state for $f_{\rm th}/2\pi$ = 5 Hz as a function of the spin-phonon coupling $g$ for different $\bar{n}_{\rm th}$. The parameters are $\tau=3.0$ ms, $\Omega/2\pi$ = 150 kHz and $\xi/2\pi$ = 0. (inset) The ratio $\frac{\delta f_{\rm th}(\bar{n}_{\rm th} \neq0)}{\delta f_{\rm th}(\bar{n}_{\rm th} =0)}$ as a function of the spin-phonon coupling $g$. (c) Parameter estimation precision in the presence of dephasing for $f_{\rm deph}/2\pi$ = 1 Hz as a function of the spin-phonon coupling $g$ for different dephasing rates $\Gamma$. (inset) The ratio $\frac{\delta f_{\rm deph}(\Gamma \neq0)}{\delta f_{\rm deph}(\Gamma =0)}$ as a function of $g$. (d) Parameter estimation precision outside the Lamb-Dicke regime for $f_{\rm LD}/2\pi$ = 1 Hz as a function of the spin-boson coupling $g$ for different value of the Lamb-Dicke parameter $\eta$. (inset) The ratio $\frac{\delta f_{\rm LD}(\eta \neq0)}{\delta f_{\rm LDR}}$ as a function of the spin-phonon coupling $g$.}
\label{fig5}
\end{figure}

The same scaling also holds as the leading term for $\xi\neq 0$, as we depict in Fig. \ref{fig2}(c).
We can generalize the result for an arbitrary odd $k$. Consider $\xi=0$ and using (\ref{alpha}) we find $\alpha=2f_{k}\bar{n}^{k/2}$ such that $\delta f_{k}\sim \bar{n}^{-k/2}$.

\section{Multi-mode nonlinearity}\label{MM}
In this section, we extend the discussion to two- and three-mode nonlinearity. Such nonlinear couplings can occur, for example, in a trapped ion system where, under the specific trap frequencies condition, the motion modes become coupled due to the mutual Coulomb repulsion. We consider an ion string with two and three ions, where two- and three-mode nonlinearity may arise.  (i) Consider an ion string with two ions, where the nonlinear interaction is given by \cite{James2003}
\begin{equation}
\hat{H}^{(ab)}_{\rm pert}=\hbar f_{ab}(\hat{a} \hat{b}^{\dag 2}+\hat{a}^{\dag}\hat{b}^{2}).\label{Hab}
\end{equation}
Here $\hat{a}^{\dag}$, $\hat{b}^{\dag}$ and $\hat{a}$, $\hat{b}$ are the creation and annihilation operators of phonon excitation in $a$-mode and $b$-mode and $\lambda=f_{ab}$ is the parameter we wish to estimate. The Hamiltonian (\ref{Hab}) describes the nonlinear processes of coherent energy exchange between two collective vibrational modes in which one phonon excitation from the $a$-mode is converted into two phonon excitations in the $b$-mode. 

(ii) Similarly, we consider the case of a three ion system, where the nonlinear Hamiltonian is \cite{James2003}
\begin{equation}
\hat{H}^{(abc)}_{\rm pert}=\hbar f_{abc}(\hat{a}^{\dag}\hat{b}\hat{c}+\hat{a}\hat{b}^{\dag}\hat{c}^{\dag}),\label{Habc}
\end{equation}
with $\hat{c}^{\dag}$ and $\hat{c}$ being the creation and annihilation operators of the phonon excitation in $c$-mode and $\lambda=f_{abc}$ is the parameter we wish to estimate. The Hamiltonian (\ref{Habc}) describes coherent exchange processes in which a phonon in the $a$-mode is converted into a phonon in the $b$-mode and a phonon in the $c$-mode.
We assume that in both (i) and (ii) cases the $a$-mode is used to implement the quantum Rabi model (\ref{model}), see Appendix \ref{PR}.

The nonlinear Ramsey interferometry is similar to the single mode case. The system is prepared initially in the state $|\psi(0)\rangle=|-\rangle|r\rangle_{a}|\psi\rangle_{b}$ for case (i) and $|\psi(0)\rangle=|-\rangle|r\rangle_{a}|\psi\rangle_{b}|\psi\rangle_{c}$ for case (ii). Here, $|\psi\rangle_{b}$, $|\psi\rangle_{c}$ are the motional states for the $b$-mode and the $c$-mode, respectively. The adiabatic evolution creates spin-motion entanglement between the spin and $a$-mode, which are decoupled from $b$-mode and $c$-mode as long as the adiabatic condition is satisfied. Note that the presence of other modes affects the adiabatic condition, since the coherent energy exchange between the vibrational modes occurs during the adiabatic evolution. However, as long as the nonlinear perturbations are weak, the adiabatic evolution creates the state
\begin{equation}
|\psi(t_f)\rangle\approx(c_{\uparrow}(t_{f})|\psi_{\uparrow,0}\rangle+c_{\downarrow}(t_{f})|\psi_{\downarrow,0}\rangle)|\psi\rangle_{b},
\end{equation}
for case (i) and, respectively,
\begin{equation}
|\psi(t_f)\rangle\approx(c_{\uparrow}(t_{f})|\psi_{\uparrow,0}\rangle+c_{\downarrow}(t_{f})|\psi_{\downarrow,0}\rangle)|\psi\rangle_{b}|\psi\rangle_{c},
\end{equation}
for case (ii). Let us assume that the motional states are coherent states $|\psi\rangle_{b}=|\alpha_{b}\rangle$ and $|\psi\rangle_{c}=|\alpha_{c}\rangle$ with  amplitudes $\alpha_{b}$ and $\alpha_{c}$. Therefore, the statistical uncertainties become
\begin{equation}
\delta f_{ab}=\frac{1}{2\pi\tau \bar{n}^{1/2}_{a}\bar{n}_{b}}\cosh(2\pi\tau f_{ab}\bar{n}_{a}^{1/2}\bar{n}_{b}) \label{errfab}
\end{equation}
and 
\begin{equation}
\delta f_{abc}=\frac{1}{2\pi\tau \bar{n}^{1/2}_{a}\bar{n}^{1/2}_{b}\bar{n}^{1/2}_{c}}\cosh(2\pi\tau f_{abc}\bar{n}^{1/2}_{a}\bar{n}^{1/2}_{b}\bar{n}^{1/2}_{c}).
\label{errfabc}
\end{equation}
Here, $\bar{n}_{b}=\langle\alpha_{b}|\hat{b}^{\dag}\hat{b}|\alpha_{b}\rangle$ and $\bar{n}_{c}=\langle\alpha_{c}|\hat{c}^{\dag}\hat{c}|\alpha_{c}\rangle$ are the average number of phonon excitations in the $b$-mode and, respectively, in the $c$-mode. In Fig. \ref{fig4} we show the statistical uncertainty for both cases. The estimation precision can be improved by increasing the initial coherent state amplitudes in $b$-mode and $c$-mode. In terms of the average number of phonon excitations, the statistical uncertainty reaches the SH limit of precision for both cases.

\section{Experimental considerations}\label{EC}
In the following section, we discuss a few sources of experimental imperfections that may reduce the estimation precision. In all of the scenarios, we consider a single mode nonlinearity (\ref{HD}) for $n=3$.
\subsection{Imperfection in the initial state preparation}
First, we consider imperfection in the initial state preparation. We assume that the initial phonon state is a thermal state described by the density matrix $\hat{\rho}_{\rm th}=\sum_{n=0}^{\infty}p_{n}|n\rangle\langle n|$, where $p_{n}=\bar{n}_{\rm th}^{n}/(1+\bar{n}_{\rm th})^{n+1}$ is the probability of observing $n$ excitations, and $\bar{n}_{\rm th}$ is the average number of initial thermal excitations. Then the initial density matrix becomes $\hat{\rho}(0)=|-\rangle\langle-|\otimes\hat{\rho}_{\rm th}$. Each of the initial components $|-\rangle\langle-|\otimes|n\rangle\langle n|$ with probability $p_{n}$ evolves adiabatically into the respective ground-state multiplet. Therefore, in the adiabatic limit we obtain:
\begin{align}
|-\rangle\langle-|\otimes|2n\rangle\langle 2n| &\rightarrow 
\langle\sigma_{z}(t_{f})\rangle = \tanh(\pi\alpha\tau), \notag\\
|-\rangle\langle-|\otimes|2n+1\rangle\langle 2n+1| &\rightarrow 
\langle\sigma_{z}(t_{f})\rangle = -\tanh(\pi\alpha\tau)
\end{align}
and the signal becomes
\begin{equation}
\langle\sigma_{z}(t_{f})\rangle=\frac{\tanh(\pi\alpha\tau)}{1+2 \bar{n}_{\rm th}}.\label{signal_th}
\end{equation}
In Fig. \ref{fig5}(a) we show the exact result for the signal compared to the expression (\ref{signal_th}), and very good agreement is observed for $\bar{n}_{\rm th}\leq 1$. For higher thermal excitations $\bar{n}_{\rm th}>1$ we observe numerical deviation from the analytical expression. We attribute this due to the stronger non-adiabatic transitions for higher number Fock states.  In Fig. \ref{fig5}(b), we plot the error propagation formula versus $g$. The result indicates that the initial thermal state reduces the estimation precision, but the same SH scaling is observed, owing to the fact that the ratio $\delta f_{\rm th}(\bar{n}_{\rm th} \neq0)/\delta f_{\rm th}(\bar{n}_{\rm th} =0)$ is approximately constant (inset of Fig. \ref{fig5}(b).  

\subsection{Spin Dephasing} 
Next, we consider the effect of spin dephasing in the system. In this regard, we use the Lindblad master equation $\partial_t \hat{\rho}=\mathcal{L}\hat{\rho}$, used to describe open systems, where $\mathcal{L}$ is the Liouvillian superoperator, which generates a completely positive map $e^{\mathcal{L}t}$ describing the time evolution of the system. For our model, the Lindblad equation reads:
\begin{equation}
\partial_{t} \hat{\rho} = -\textit{i}[\hat{H}(t), \hat{\rho}] + \frac{\Gamma}{2} \mathcal{D}_{\hat{\sigma}_{z}}[\hat{\rho}],
\end{equation}
where $\Gamma=1/\tau_{\rm dec}$ is the dephasing rate, $\tau_{\rm dec}$ is the decoherence time, and $\mathcal{D}_{\hat{\sigma}_{z}}[\hat{\rho}]=\sigma_z\hat{\rho}\sigma_z-\hat{\rho}$ is the Lindblad superoperator with jump operator $\sigma_z$. In Fig. (\ref{fig5})(c) we show the statistical uncertainty for different dephasing rates $\Gamma$. Decreasing the decoherence time $\tau_{\rm dec}$ leads to a decrease in the estimation precision, but, for small dephasing rates, the SH scaling is preserved, as seen from the ratio $\delta f_{\rm deph}(\Gamma \neq0)/\delta f_{\rm deph}(\Gamma =0)$ in the inset, which is approximately constant for $\Gamma/2\pi=1$ Hz and $\Gamma/2\pi=5$ Hz. We note that, by encoding the qubit in a magnetic-field-insensitive hyperfine transition, one can significantly reduce the effect of spin dephasing, as experimentally demonstrated in \cite{Langer2005}. 

\subsection{Outside the Lamb-Dicke regime} 
Finally, we consider working outside the Lamb-Dicke regime. The regime is quantitatively expressed as $\eta\langle(a^{\dagger}+a)^{2}\rangle^{1/2} \ll 1$, where $\eta$ is the Lamb-Dicke parameter. In order to investigate the system outside of this regime, we consider a nonlinear coupling term between the spin and the motion mode of the form:
\begin{equation}
\hat{H}_{\rm coup} =  \hbar g \sigma_z (\hat{a}^{\dagger}\hat{F}(\hat{n})+\hat{F}(\hat{n})\hat{a}), \label{LD}
\end{equation}
which describes the interaction between the spin and the phonon mode outside the Lamb-Dicke regime. The nonlinear operator $\hat{F}(\hat{n})$ is defined as \cite{Vogel1995}:
\begin{equation}
\hat{F}(\hat{n}) = e^{-\eta^2/2}\sum_{n=0}^{\infty}\frac{(-\eta^2)^n}{n!(n+1)!}\hat{a}^{\dagger n}\hat{a}^{n}.
\end{equation}
In Fig. \ref{fig5}(d) we display the error propagation formula and see that up to $g/2\pi\sim7.5$ kHZ the SH is preserved as in the previous two cases. Surprisingly, for $g/2\pi>9$ kHz we even see an improvement in the estimation precision for the cases of $\eta=0.2$ and $\eta=0.3$ over the Lamb-Dicke regime case, as seen from the ratio $\delta f_{\rm LD}(\eta \neq0)/\delta f_{\rm LDR}$ in the inset.

\section{Conclusion}\label{C}
We have discussed an adiabatic Ramsey interferometry for measuring weak nonlinearity with trapped ions. Such nonlinearities may arise due to anharmonic terms in the trapping potential of a single ion or due to higher-order terms in the Coulomb interaction expansion which causes coupling between the motion modes. The essence of the method is the adiabatic creation of the Schr\"odinger cat state with probability amplitudes depending on the unknown parameter. This allows us to estimate the nonlinear coupling by detection of the spin probabilities alone after the adiabatic evolution. We have shown that the sensitivity is improved by the mean phonon excitations and reaches the SH limit of precision.

Our sensing technique does not require specific initial entangled state preparation and can even be applied to an initial thermal motion state. We have shown that the SH scaling can be observed even in the presence of weak spin-dephasing and can be applied outside the Lamb-Dicke approximation. Furthermore, our sensing technique is also relevant for other quantum-optical platforms such as cavity and circuit QED.





\section*{Acknowledgments}
We acknowledge the Bulgarian national plan for recovery and resilience, contract BG-RRP-2.004-0008-C01 (SUMMIT: Sofia University Marking Momentum for Innovation and Technological Transfer), project number 3.1.4.

	\begin{appendix}
		\section{Physical Realization}\label{PR}
		
	Consider a chain of two ions. Including the third-order term arising from the Coulomb repulsion between the ions, the motion Hamiltonian is:
    \begin{equation}
    \hat{H}_{\rm v}=\hbar\omega_{a}\hat{a}^{\dag}\hat{a}+\hbar\omega_{b}\hat{b}^{\dag}\hat{b}+\hbar f_{ab}(\hat{a}\hat{b}^{\dag 2}+\hat{a}^{\dag}\hat{b}^{2}).
    \end{equation}
Here $\hat{a}^{\dag}$, $\hat{a}$ are the creation and annihilation operators of phonon excitation in the axial breathing mode with frequency $\omega_{a}$ and respectively, $\hat{b}^{\dag}$, $\hat{b}$ correspond to the radial rocking mode with frequency $\omega_{b}$. In order to ensure the survival of the nonlinear term under the rotating-wave-approximation, we impose the resonance condition $\omega_{a}\approx2\omega_{b}$.

We assume that the first ion interacts with bichromatic laser fields propagating along the trap axis, with frequencies $\omega_{R}=\omega_{0}-\omega_{a}+\omega$ and $\omega_{B}=\omega_{0}+\omega_{a}-\omega$ \cite{Wineland1998,Schaetz2012,Haffner}. Additionally, the qubit states are driven by a resonant laser field with Rabi frequency $\Omega(t)$, and the trap frequencies satisfy the condition, $\omega_{a}=2\omega_{b}+\omega$. Moving in the rotating frame with respect to $\hat{U}_{R}(t)=e^{-\omega_{0}t\sigma^{z}_{1}-i(\omega_{a}-\omega)t \hat{a}^{\dag}\hat{a}-i\omega_{b}t\hat{b}^{\dag}\hat{b}}$ and assuming Lamb-Dicke approximation, the interaction Hamiltonian becomes
\begin{eqnarray}
\hat{H}(t)&=&\hbar\omega \hat{a}^{\dag}\hat{a}+\frac{\hbar\Omega(t)}{2}\sigma_{y}+
\hbar g\sigma_{x}(\hat{a}^{\dag}+\hat{a})\notag\\
&&+\hbar f_{ab}(\hat{a}\hat{b}^{\dag 2}+\hat{a}^{\dag}\hat{b}^{2}).
\end{eqnarray}
The phonon squeezing term can be realized, for example, by modulating the confining potential or by applying an additional time-varying electric field. 

Similarly, we consider a linear ion crystal with three ions. Then the motional Hamiltonian becomes
\begin{equation}
\hat{H}_{\rm v}=\hbar\omega_{a}\hat{a}^{\dag}\hat{a}+\hbar\omega_{b}\hat{b}^{\dag}\hat{b}+\omega_{c}\hat{c}^{\dag}\hat{c}+
\hbar f_{abc}(\hat{a}^{\dag}\hat{b}\hat{c}+\hat{a}\hat{b}^{\dag}\hat{c}^{\dag}).
\end{equation}
Here, one phonon from the axial zigzag mode ($a$-mode) is converted into one phonon in the radial rocking mode ($b$-mode) and one in the radial zigzag mode ($c$-mode). The realization of the quantum Rabi model is similar as above but now the trap frequency condition is $\omega_{a}=\omega_{b}+\omega_{c}+\omega$.	

Finally, the first-order nonlinear term that arises due to the anharmonicity of the trap potential is $\hat{V}(z)=\kappa f(t)\hat{z}^{3}$ with $\hat{z}=z_{0}(\hat{a}^{\dag}+\hat{a})$ \cite{Home2011}. Here, we assume that the axial trap potential is additionally time-modulated through $f(t)=\cos^{2}(\omega_{d}t)$, with driving frequency $\omega_d$. Setting the condition $\omega_{d}=3/2(\omega_{z}-\omega)$ and applying rotating-wave approximation, we have $\hat{H}_{\rm v}=\omega \hat{a}^{\dag}\hat{a}+f_{3}(\hat{a}^{\dag 3}+\hat{a}^{3})$, where $f_{3}=\kappa z^{3}_{0}/4$.

\end{appendix}




\appendix

\end{document}